**Diffusion of exciplex.**

**I. Energy transfer from exciplex to exciplex-forming pair**

Hwang-Beom Kim,[1] and Jang-Joo Kim[1,2,*]

[1]*Department of Materials Science and Engineering and the Center for Organic Light Emitting Diodes, Seoul National University, Seoul 151-742, South Korea.*

[2]*Research Institute of Advanced Materials (RIAM), Seoul National University, Seoul 151-744, South Korea.*

**Abstract**

Exciton diffusion in organic films is crucial phenomenon in optoelectronic devices such as organic light-emitting diodes and organic photovoltaics. However, diffusion of exciplexes has not been actively studied despite their ever-growing importance in such devices due to the lack of apparent charge-transfer absorption, resulting in the absence of energy transfer (ET) to exciplex-forming pairs. Here, the ET from exciplexes to exciplex-forming pairs is reported by analysis of transient photo-luminescent profiles. Recent reports and our own observation of the sub-bandgap charge-transfer absorption in exciplex-forming systems support the ET from exciplexes to exciplex-forming pairs. The ET mechanism is discussed in the following paper.

**1. Introduction**

Excited-state charge-transfer complexes (exciplexes) are formed by charge transfer (CT) in the excited state between electron donor molecules and nearby electron acceptor molecules after the photo-excitation of the electron donor or acceptor molecules. Exciplexes are widely observed in nature, such as photosynthetic systems, and also organic optoelectronic devices such as organic photovoltaics (OPVs) or organic light-emitting diodes (OLEDs), where hetero-junctions between electron donor and acceptor molecules are employed. Recently, exciplexes have been actively used as thermally assisted delayed fluorescence emitters in OLEDs to take advantage of their higher luminescent exciton ratio, due to their low energy difference between the singlet and triplet excited states [1–4]. Exciplexes have also been used as host materials in highly efficient OLEDs, exploiting low driving voltage, good charge balance, and low efficiency roll-off[5–7].

Exciton diffusion in organic films is crucial phenomenon in such optoelectronic devices. Its mechanism is Förster-type resonance energy transfer (FRET) and Dexter-type exchange energy transfer (DET). FRET is dominant mechanism for singlet exciton diffusion. On the other hand, DET is the only mechanism for triplet exciton diffusion of non-phosphorescent molecules because triplet states generally have the very low oscillator strength [8]. However, the diffusion of exciplexes has not been studied until very recently, primarily because of the lack of apparent CT absorption in exciplex-forming systems measured by standard steady-state absorption experiments, which is a necessary condition for diffusion to take place via energy transfer (ET) mechanisms. The lack of apparent CT absorption in standard steady-state absorption experiments in exciplex-forming systems is the reason for the term "exciplex".

Here, we report that ET from exciplexes to exciplex-forming pairs by analysis of the transient analysis of high-energy exciplexes and low-energy exciplexes when they are close. Recent reports and our own observation for the sub-bandgap CT absorption in organic donor-acceptor systems support the ET from exciplexes to exciplex-forming pairs [9–14].

## 2. Experiments

Organic materials used in this paper were purchased from Nichem Fine Technology. Thermal deposition on precleaned quartz substrates was used to fabricate all organic films at a base pressure of $< 5 \times 10^{-7}$ Torr. All films were deposited at a rate of 1 Å s$^{-1}$. Extinction coefficients were measured using variable angle spectroscopic ellipsometry (VASE, J. A. Woolam M-2000 spectroscopic ellipsometer). A spectrofluorometer (Photon Technology International, Inc.) with an incorporated monochromator (Acton Research Co.) was used for PL spectra measurements, and a streak camera (C10627, Hamamatsu Photonics) excited by a nitrogen gas laser with a pulse width of 800 ps (KEN-2X, Usho) was used for transient and time-resolved PL experiments.

## 3. Results

We conducted two different experiments to investigate the ET from exciplexes to exciplex-forming pairs. The first experiment exploited a bilayer system with two different kinds of exciplexes. The second one was carried out for a co-doped system with two different kinds of exciplexes.

Firstly, an exciplex-forming TAPC:PO-T2T mixed film (1:1 molar ratio, 2 nm thick) was stacked on another exciplex-forming mCBP:PO-T2T mixed film (1:1 molar ratio, 5 nm thick) (Film A in Fig. 1(a)), and the emission characteristics of the bilayer film were compared with those of non-stacked single-layer films of the same thickness, where TAPC, PO-T2T, and mCBP represent di-[4-(*N*,*N*-ditolyl-amino)-phenyl]cyclohexane, (1,3,5-triazine-2,4,6-triyl)tris(benzene-3,1-diyl))tris(diphenylphosphine oxide), and 4,4′-bis(3-methylcarbazol-9-yl)-2,2′-biphenyl, respectively, whose molecular structures are shown Fig. 1(a). Formation of exciplexes in the mCBP:PO-T2T and TAPC:PO-T2T films was confirmed by the featureless redshifted emission from the constituent materials (Fig. S1(a)). mCBP and TAPC were used as electron donors and PO-T2T was used as an electron acceptor for the exciplexes. The thickness of the films was adjusted for clear observation of the interface quenching effect (or ET) in the transient PL experiments. Fig. 1(b) compares the emission spectra for 9 μs after excitation of the single-layer films and the bilayer film when the films were excited with a $N_2$ pulsed laser at the wavelength of 337 nm. The emission from the bilayer film is composed of the exciplex emissions from the consisting layers. The longer wavelength emission from the TAPC:PO-T2T layer with the peak wavelength of 560 nm is significantly increased and the exciplex emission from the mCBP:PO-T2T layer with the peak wavelength of 475 nm was reduced significantly in Film A compared to the single layer emissions. The energy levels of the mCBP, TAPC, and PO-T2T excitons, and the mCBP:PO-T2T and TAPC:PO-T2T exciplexes are shown in Fig. 1(c). The energy levels of the mCBP, TAPC, and PO-T2T singlet excitons were estimated from the onset of their absorption spectra. The energy levels of the mCBP, TAPC, and PO-T2T triplet excitons and mCBP:PO-T2T and TAPC:PO-T2T singlet exciplexes were estimated from the onset of their integrated emission spectra. The triplet energy levels of exciplexes would be similar to the singlet levels due to small overlap between the highest occupied molecular orbital and the lowest unoccupied molecular orbital. The normalized transient PL intensities from the single layers are compared to those of the bilayer for the mCBP:PO-T2T exciplex emission (detection wavelength 455–475 nm) and the TAPC:PO-T2T exciplex emission (detection wavelength 620–660 nm) in Fig. 1(d). The results show that the lifetime of the delayed emission of the mCBP:PO-T2T exciplex in the stacked layer is significantly reduced compared to that of the single layer. In contrast, the lifetime of delayed emission from the TAPC:PO-T2T exciplex in the stacked layer is significantly increased compared to that of the single layer. The transient results show that ET takes place from the mCBP:PO-T2T exciplex to the TAPC:PO-T2T exciplex-forming pair.

Similar behavior was observed in a different exciplex-forming system where two electron donor molecules [4,4′,4′′-tri(*N*-carbazolyl)triphenylamine (TCTA) and 4,4′,4′′-tris (3-methyl-phenylphenylamino) triphenylamine (m-MTDATA)] were co-doped in an electron acceptor [PO-T2T] at 5 mol.% each (Film B in Fig. 2(a)). The molecular structures of TCTA and m-MTDATA are shown in Fig. 2(a). The featureless redshifted emission spectra of the mixed films of TCTA:PO-T2T and m-MTDATA:PO-T2T in Fig. S1(b) confirm the formation of exciplexes between TCTA or m-MTDATA and PO-T2T. The extinction coefficients of TCTA, m-MTDATA, and PO-T2T neat films and m-MTDATA:PO-T2T (m.r 1:1) mixed film [15] are shown in Fig. 2(b) along with the integrated emission spectra of a TCTA:PO-T2T (m.r. 5:95) doped film, an m-MTDATA:PO-T2T (m.r. 5:95) doped film, and Film B for 500 ns after excitation when the electron donors in the films were selectively excited by a $N_2$ pulsed laser at the wavelength of 337 nm. Again the emission from Film B is composed of the emissions from the TCTA:PO-T2T and m-MTDATA:PO-T2T exciplexes, which are the high-energy exciplexes and low-energy exciplexes, respectively. The energy levels of the system are shown in Fig. 2(c), and were estimated using the same methods as those in Fig. 1(c). The emission from Film B is composed of the emissions from the TCTA:PO-T2T and m-MTDATA:PO-T2T exciplexes as expected. However, the intensity of the TCTA:PO-T2T exciplex in Film B is significantly reduced to 23% compared to the singly doped TCTA:PO-T2T (m.r. 5:95) film and the m-MTDATA:PO-T2T exciplex emission increased by the similar ratio. This fact indicates that the energy is transferred from the excited states involving TCTA molecules to those involving m-MTDATA molecules considering the energy levels of the singlet excitons of TCTA and m-MTDATA, and the singlet exciplexes of the TCTA:PO-T2T and m-MTDATA:PO-T2T. Fig. 2(d) compares the transient PL profiles of the TCTA:PO-T2T exciplex emission from Film B to that from the singly doped TCTA:PO-T2T film measured in the wavelength range 510 to 530 nm, where the TCTA:PO-T2T exciplex emission is dominant. The decay rate constant for the emission of the TCTA:PO-T2T exciplexes in Film B was larger than that in the singly doped TCTA:PO-T2T film. On the other hand, the lifetime of the m-MTDATA:PO-T2T exciplexes in Film B was longer than that in the singly doped m-MTDATA:PO-T2T film in the detection-wavelength range of 700 to 800 nm, where m-MTDATA:PO-T2T exciplex emission is dominant.

The extinction coefficients of a m-MTDATA:PO-T2T (m.r 1:1) mixed film are shown in Fig. 2(b) along with the extinction coefficients of TCTA, m-MTDATA, and PO-T2T neat films. The exciplex-forming mixed film showed clear absorption in the sub-bandgap region at the longer wavelength than 420 nm even though it is very weak. Thick films were used for the purpose of detecting the intermolecular CT absorption by a simple absorption equipment of reflection/transmission measurement. Then, absorbance for the exciplex-forming mixed film in the sub-bandgap region was analyzed using the transfer matrix method to obtain extinction coefficient for the intermolecular CT state as low as $10^{-4}$ [15].

## 4. Discussion

The reduction of the lifetime and the intensity of the high-energy exciplex PL and their increase of the low-energy exciplex PL in the bilayer structure (Fig. 1) and in the doped layer (Fig. 2) clearly indicate that the ET from exciplexes to exciplex-forming pairs takes place. This is the first observation of the possibility of ET from an exciplex to an exciplex-forming pair to our best knowledge, resulting in the diffusion of exciplexes. The existence of the sub-bandgap CT absorption in the exciplex-forming film (Fig. 2(b) bottom) supports the possibility of the ET. Exciplex dissociation, followed by Langevin recombination, can be considered as an origin for the transient behaviors of the exciplexes. The mCBP:PO-T2T exciplexes and TCTA:PO-T2T exciplexes which are the high-energy exciplexes in each system could be dissociated into the free polarons, and they can recombine into the low-energy exciplexes in the organic films. Since Langevin recombination is a bimolecular process, the transient behavior of the exciplex emission must depend on the density of the charged species at the moment of photo-excitation. The initial numbers of free electrons and holes are both proportional to the intensity of the excitation light. The square dependence of exciplex generation rate will result in the intensity dependence of decay rates if Langevin recombination is dominant process for the generation of the singlet exciplex states. The transient PL profiles for the mCBP:PO-T2T film (m.r 1:1) and TCTA:PO-T2T film (m.r 5:95) where the high-energy exciplexes for each system only exist exhibit negligible change with different excitation light intensities, as shown in Fig. 3. The excitation light intensities were controlled by ND filters with different optical densities ranging from 25 to 800 µW. These results indicate that bimolecular recombination or Langevin recombination hardly participate in the decay of the exciplexes to ground states.

To ensure that the ET from the exciplex to the exciplex-forming pair in Film B can describe their transient decay profiles quantitatively, we carried out the fitting of the transient PL profiles for the TCTA:m-MTDATA:PO-T2T mixed system as shown in Fig. 2(d). The transient PL profiles for the TCTA:PO-T2T exciplex were fitted by a bi-exponential decay model considering the delayed fluorescence of the exciplexes as follows [2];

$$I_{\text{norm.PL}} = I_{\text{PL}}^{\text{prompt}} + I_{\text{PL}}^{\text{delayed}} = \exp(-k_{\text{prompt}}t) + C_{\text{delayed}}\left(\exp(-k_{\text{delayed}}t) - \exp(-k_{\text{prompt}}t)\right), \quad (1)$$

where $I_{\text{norm.PL}}$, $I_{\text{PL}}^{\text{prompt}}$, and $I_{\text{PL}}^{\text{delayed}}$ are the normalized transient PL intensity, prompt transient PL intensity, and delayed transient PL intensity from exciplexes, respectively, $k_{\text{prompt}}$ and $k_{\text{delayed}}$ are the prompt and delayed decay rate constants, respectively, $C_{\text{delayed}}$ is the pre-exponential constant for the delayed decay, and $t$ is the time after excitation. The prompt transient PL intensity is the single-exponential term due to the decay of the singlet excited state into the ground state and triplet exited state, corresponding to the first term in the right-hand side of Eq. (1). The delayed decay of the exciplex originates from the intersystem crossing from the singlet excited state to the triplet excited state, followed by the reverse intersystem crossing from the triplet state to the singlet state, of which intensity corresponds to the second term in the right-hand side of Eq. (1). 1 - $C_{\text{delayed}}$ is the pre-exponential constant for the prompt decay ($C_{\text{prompt}}$) for the normalized transient PL intensity. The ET rate constant from the TCTA:PO-T2T exciplexes to the m-MTDATA:PO-T2T exciplex-forming pairs, $k_{\text{ET}}$, was calculated as the difference between the prompt decay rate constants of TCTA:PO-T2T exciplexes in the TCTA:PO-T2T film and Film B [2], which is $9.72 \times 10^6$ s$^{-1}$.

We fitted the transient PL profiles for the m-MTDATA:PO-T2T exciplex in Film B based on the experimental ET rate constant ($9.72 \times 10^6$ s$^{-1}$) for the singlet exciplex states. The differential rate equations for the prompt part of transient PL profiles in Film B are as follows.

$$\frac{d[(T:P)^*]}{dt} = -k_{T:P}^{prompt}[(T:P)^*] - k_{ET}[(T:P)^*]$$
$$= -2.84 \times 10^7 [(T:P)^*] - 9.72 \times 10^6 [(T:P)^*] \quad (2)$$

$$\frac{d[(m:P)^*]}{dt} = -k_{m:P}^{prompt}[(m:P)^*] + k_{ET}[(T:P)^*]$$
$$= -5.80 \times 10^7 [(m:P)^*] + 9.72 \times 10^6 [(T:P)^*] \quad (3)$$

where $[(T:P)^*]$ and $[(m:P)^*]$ are the concentrations of the singlet TCTA:PO-T2T and m-MTDATA:PO-T2T exciplexes in Film B, respectively, $k_{T:P}^{prompt}$ and $k_{ET}$ are the prompt decay rate constant of the TCTA:PO-T2T exciplex in the TCTA:PO-T2T film and the ET rate constant from the singlet TCTA:PO-T2T exciplex to the m-MTDATA:PO-T2T exciplex-forming pair, respectively, and $k_{m:P}^{prompt}$ is the prompt decay rate constant of the m-MTDATA:PO-T2T exciplex in the m-MTDATA:PO-T2T film which is $5.80 \times 10^7$ (in units of s$^{-1}$) from the two exponential fitting of the corresponding transient PL profile as shown in Fig. 2(d) (ii) (black solid line). We did not consider ET in the delayed region for the TCTA:m-MTDATA:PO-T2T system because the ET from TCTA:PO-T2T triplet exciplex to m-MTDATA molecules as well as to m-MTDATA:PO-T2T exciplex-forming pairs can occur considering their energy levels as shown in Fig. 2(c). This will complicate the analysis of the ET rate constant of TCTA:PO-T2T triplet exciplexes to m-MTDATA:PO-T2T exciplex-forming pairs from the transient PL profiles. The initial concentration ratio of the TCTA:PO-T2T exciplexes and m-MTDATA:PO-T2T exciplexes for the rate equations was set to be 2:5 considering the extinction coefficients of TCTA and m-MTDATA at excitation wavelength of 337 nm, and the ET from TCTA singlet excitons to m-MTDATA before TCTA singlet excitons form the TCTA:PO-T2T singlet exciplexes. Note that the more initial excited energy acceptors increase the influence of decay rate constant of the intrinsic energy acceptor on the transient PL profiles.

The solutions of the differential rate equations are

$$I_{\text{T:P}}^{\text{prompt}} = \exp(-3.81 \times 10^7 t) \tag{4}$$

$$I_{\text{m:P}}^{\text{prompt}} = 0.81\exp(-5.80 \times 10^7 t) + 0.19\exp(-3.81 \times 10^7 t). \tag{5}$$

They are plotted in Fig. 2(d) as red dotted lines (prompt part). The prompt part of the TCTA:PO-T2T exciplexes in Film B is a single-exponential line with a decay time constant of 26 ns, which is the reciprocal of the decay rate constant, and that of the m-MTDATA:PO-T2T exciplexes in Film B is a two-exponential line with the decay time constants of 17 ns and 26 ns. Because of the larger number of the m-MTDATA:PO-T2T exciplexes than the TCTA:PO-T2T exciplexes in Film B at the excitation moment, the pre-exponential constant for the decay time constant of 26 ns is 0.19 leading to small change of the transient PL profiles for the m-MTDATA:PO-T2T exciplexes than those for the TCTA:PO-T2T exciplexes. From the tail-fitting for the delayed part of the m-MTDATA:PO-T2T exciplex in Film B, the fit line of the entire transient PL profile for the m-MTDATA:PO-T2T exciplex in Film B well matches with the experimental values as shown in Fig. 2(d). The fitting parameters for the fit lines (solid lines) for four kinds of the normalized transient PL profiles in Fig. 2(d) are given in Table 1. The delayed part (dashed line) corresponds to the difference between the fit line (solid line) and the prompt part (dotted line) for four kinds of the transient PL profiles. Very good fittings confirm that ET takes place indeed from the TCTA:PO-T2T exciplex to the m-MTDATA:PO-T2T exciplex-forming pair.

## 5. Conclusion

It is demonstrated that ETs from exciplexes to exciplex-forming pairs take place based on observations of exciplex quenching by other kinds of exciplex-forming pairs in the transient analysis. The consistency of the transient PL profiles with various excitation intensities of the exciplexes indicates that the energy migration via bimolecular recombination hardly takes place. The ET from exciplexes to exciplex-forming pairs is supported by the observation for the sub-bandgap CT absorption in exciplex-forming films because the spectral overlap between absorption and emission spectra is necessary for the ET process.

**Acknowledgements**

This work was supported by a grant from National Research Foundation (NRF) funded by the Ministry of Science and ICT (MSIT).

**Table 1.** Fitting parameters for the normalized transient PL profiles of the TCTA:PO-T2T exciplexes in the TCTA:PO-T2T film and Film B and those of the m-MTDATA:PO-T2T exciplexes in the m-MTDATA:PO-T2T film and Film B.

| Detection wavelength | Film | $C_{prompt}$ | $k_{prompt}$ (s$^{-1}$) | $k_{delayed}$ (s$^{-1}$) |
|---|---|---|---|---|
| 510-530 nm | TCTA:PO-T2T | 0.77 | 2.84×10$^7$ | 1.09×10$^6$ |
| | Film B | 0.85 | 3.81×10$^7$ | 2.36×10$^6$ |
| 700-800 nm | m-MTDATA:PO-T2T | 0.89 | 5.80×10$^7$ | 2.85×10$^6$ |
| | Film B | 0.70 | 5.80×10$^7$ | 2.56×10$^6$ |
| | | 0.16 | 3.81×10$^7$ | |

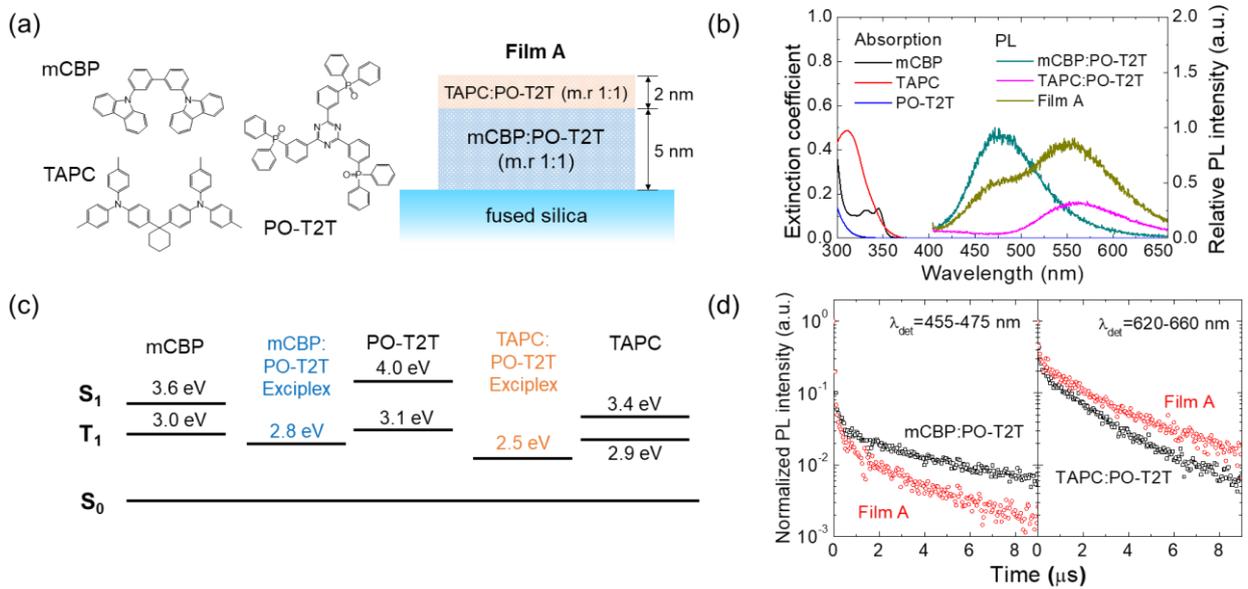

FIG. 1. (a) Molecular structures of mCBP, TAPC, and PO-T2T and a bilayer structure of Film A. (b) Extinction coefficients of mCBP, TAPC, and PO-T2T films and PL spectra of an mCBP:PO-T2T film, a TAPC:PO-T2T film, and Film A integrated for 9 μs after excitation at 337 nm. (c) Energy levels of mCBP, TAPC, and PO-T2T excitons and mCBP:PO-T2T and TAPC:PO-T2T exciplexes, where $S_1$ and $T_1$ represent the lowest excited singlet and triplet excited states. (d) Normalized transient PL intensities for the mCBP:PO-T2T film (m.r 1:1) and Film A in the wavelength range of 455 to 475 nm (left) and those for the TAPC:PO-T2T (m.r 1:1) film and Film A in the wavelength range of 620 to 660 nm (right).

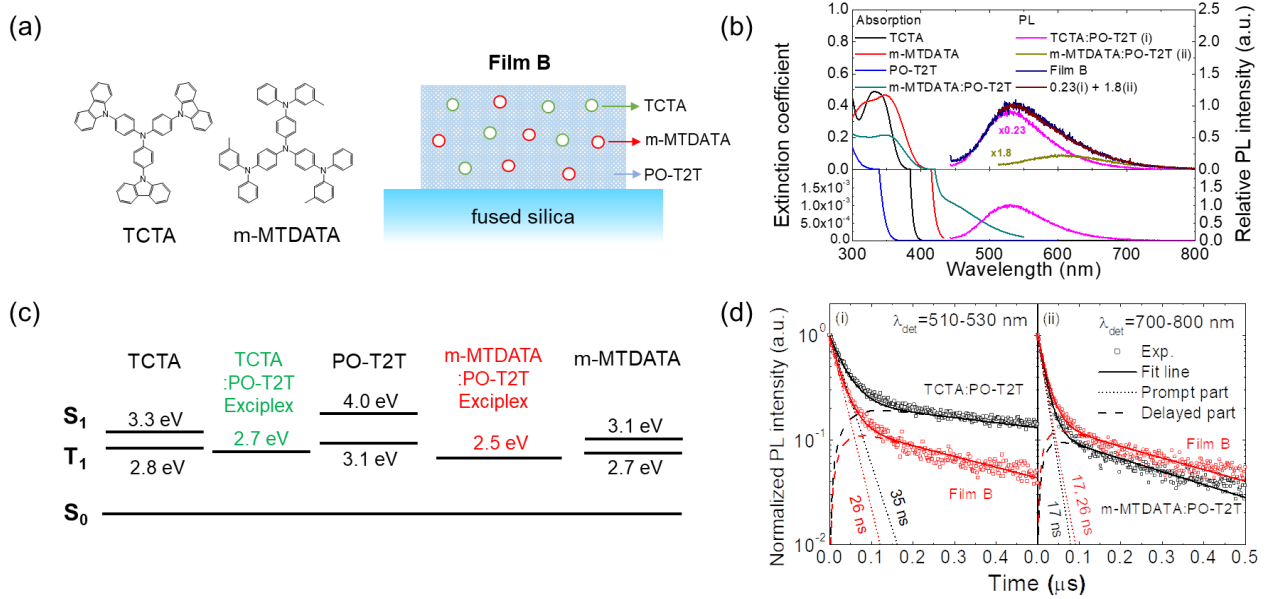

FIG. 2. (a) Molecular structures of TCTA and m-MTDATA, and a PO-T2T film co-doped with TCTA and m-MTDATA (Film B). (b) Extinction coefficients of TCTA, m-MTDATA, and PO-T2T films and m-MTDATA:PO-T2T (m.r 1:1) film and integrated PL spectra for 500 ns after excitation of aTCTA:PO-T2T exciplex (m.r. 5:95) film (i) (×0.23), an m-MTDATA:PO-T2T (m.r. 5:95) film (ii) (×1.8), and Film B (×1) at 337-nm excitation, and the sum of 0.23(i) and 1.8(ii) (upper part), and the extinction coefficients below 0.002 and the PL spectra (lower part). (c) Energy levels of $S_1$ and $T_1$ of TCTA, m-MTDATA and PO-T2T excitons, and TCTA:PO-T2T and m-MTDATA:PO-T2T exciplexes. (d) Normalized transient PL intensities (empty squares), fit lines of the total (solid lines), the prompt (dotted line) and delayed (dashed line) parts for the TCTA:PO-T2T film (m.r 5:95) and Film B in the wavelength range of 510 to 530 nm (left) and those of the m-MTDATA:PO-T2T (m.r 5:95) film and Film B in the wavelength range of 700 to 800 nm (right). The decay time constants for the prompt parts are also given.

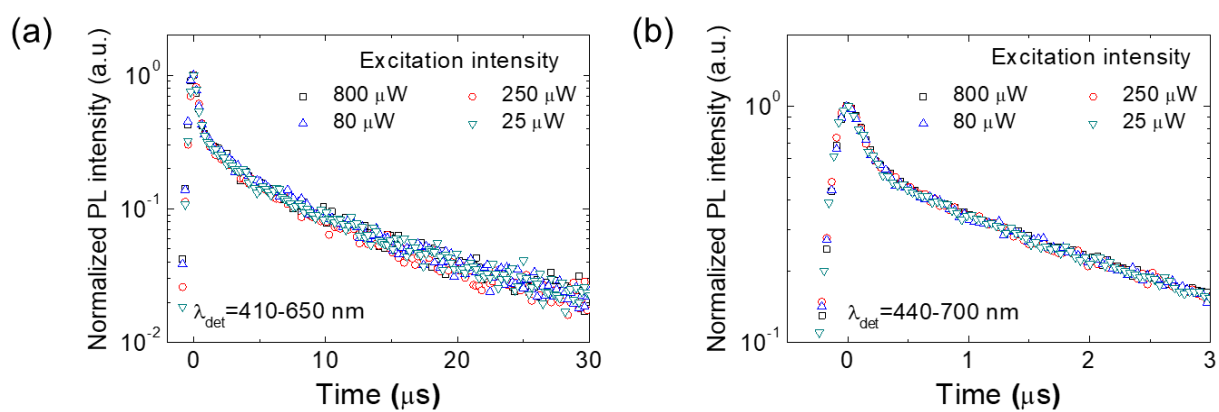

FIG. 3. (a) Normalized transient PL profiles of an mCBP:PO-T2T film (m.r 1:1) excited by a 337 nm pulsed laser light (N$_2$ laser) at different intensities in the wavelength range of 390 to 650 nm. (b) Normalized transient PL profiles of a TCTA:PO-T2T film (m.r 5:95) at different intensities in the wavelength range of 440 to 700 nm.